\documentclass[9pt]{Interspeech}

\usepackage{graphicx}
\usepackage{svg}
\usepackage{array}
\usepackage{tabularx}
\usepackage{cite}
\usepackage{amsmath,amssymb,amsfonts}
\usepackage{algorithmic}
\usepackage{graphicx}
\usepackage{textcomp}
\usepackage{xcolor}
\usepackage{arydshln}
\usepackage{boldline, makecell, hhline, pifont}
\usepackage{booktabs}

\interspeechcameraready

\usepackage{comment}

\title{Speech Enhancement based on cascaded two flows }
\begin{document}

\author[affiliation={}]{Seonggyu}{Lee}
\author[affiliation={}]{Sein}{Cheong}
\author[affiliation={}]{Sangwook}{Han}
\author[affiliation={}]{Kihyuk}{Kim}
\author[affiliation={}]{Jong Won}{Shin}



\affiliation[nocounter]{}{Gwangju Institute of Science and Technology}{Korea}
\email{\{lsqjin2022,seiinjung,swhan9873, kpaul073\}@gm.gist.ac.kr, jwshin@gist.ac.kr }

\newcommand{\blue}[1]{\textcolor{blue}{#1}}
\maketitle
\begin{abstract}
Speech enhancement (SE) based on diffusion probabilistic models has exhibited impressive performance, while requiring a relatively high number of function evaluations (NFE). 
Recently, SE based on flow matching has been proposed, which showed competitive performance with a small NFE. 
Early approaches adopted the noisy speech as the only conditioning variable.
There have been other approaches which utilize speech enhanced with a predictive model as another conditioning variable and to sample an initial value, but they require a separate predictive model on top of the generative SE model. 
In this work, we propose to employ an identical model based on flow matching for both SE and generating enhanced speech used as an initial starting point and a conditioning variable. 
Experimental results showed that the proposed method required the same or fewer NFEs even with two cascaded generative methods while achieving equivalent or better performances to the previous baselines.\textsuperscript{†}
\renewcommand{\thefootnote}{†}
\footnotetext{Our codes are available at online :\\ \url{https://github.com/seongq/cascadingtwoflowmatching}}
\end{abstract}

\section{Introduction}
Speech enhancement (SE) aims to restore clean speech signals from those contaminated by environmental noises \cite{loizou2007speech, minseungenhancement,seinenhancement, enhancementwangtan,  kim2022factorizedfmvdr,kim2021target, enhancementgan,enhancementflow, enhancementvae,sgmsep,refgerkmannenhancement1,vpidm, Lay2023bbed,  lemercier2023storm, thunderinterspeech2024}. 
Traditional methods often utilize the statistical characteristics of clean speech signals and environmental noises \cite{minseungenhancement,seinenhancement}, but most of  recent studies employ deep neural networks (DNNs) to estimate clean speech signals \cite{enhancementwangtan,  kim2022factorizedfmvdr, kim2021target}. 
Generative approaches that focus on modeling the underlying distribution of clean speech signals have recently been introduced \cite{enhancementgan,enhancementflow, enhancementvae,sgmsep,refgerkmannenhancement1,vpidm, Lay2023bbed,  lemercier2023storm, thunderinterspeech2024}. 
Among them, the SE based on diffusion probabilistic models which utilizes stochastic differential equations (SDEs) have demonstrated remarkable performance \cite{sgmsep,refgerkmannenhancement1,vpidm, Lay2023bbed,  lemercier2023storm,thunderinterspeech2024}. 
This class of approaches requires repeated evaluation of a DNN model that approximates the score function to estimate clean speech. 
The number of times that the DNN model is evaluated, called the number of function evaluations (NFE), typically exceeds 25 \cite{sgmsep,refgerkmannenhancement1,vpidm, Lay2023bbed,  lemercier2023storm}, which may limit the applicability of the diffusion model-based SE.  

Flow matching (FM) to model continuous normalizing flows (CNFs) which converts a random vector following a simple distribution into another one with a complex distribution through invertible transformations has been proposed as an alternative to the diffusion probabilistic models \cite{lipmanflow, tong2024improving}. 
The FM with a conditional flow matching (CFM) loss and the optimal transport (OT) conditional vector field showed faster sampling and better performance than the previous diffusion models in several tasks \cite{lipmanflow, tong2024improving}. 
It has also been applied to speech processing such as speech separation \cite{speechflow}, speech enhancement \cite{speechflow, lee2025flowse}, and audio-visual speech enhancement \cite{flowavse}. 
Among them, the FlowSE in \cite{lee2025flowse} adopts the noisy speech as an additional condition and modifies the OT conditional vector field so that the mean of the conditional probability path moves linearly from noisy speech to clean speech and the standard deviation decreases linearly. 
It showed equivalent or better performance to the previous diffusion model-based SE with a fewer NFE.

The DNN models that estimate scores in the diffusion model-based SE or vector fields in the flow matching-based SE have conditioning variables other than the state and noise level at a specific time. 
Noisy speech is given to the model in \cite{sgmsep, Lay2023bbed,vpidm, refgerkmannenhancement1, flowavse, speechflow, lee2025flowse}, while speech enhanced with an additional predictive SE model is used as a conditioning variable along with noisy speech in \cite{lemercier2023storm, tai2023revisiting,serra2022universal, kim2024guided}. 
The utilization of the enhanced speech brought about performance improvement, but it requires a separate predictive SE model. 

In this paper, we propose a generative model that cascades two flows for SE where two flows are approximated by a single model. 
The first flow models the transformation from a random vector following a simple distribution centered on noisy speech to the one distributed by the probability distribution of clean speech given noisy speech just like FlowSE. 
The output of the first flow is used as an additional conditioning variable and the mean of the starting point for the second flow. 
Experimental results showed that the proposed method outperformed previously proposed approaches without increasing the total NFE. 
\section{Related Works}
\subsection{Diffusion model-based Speech Enhancement}
In the diffusion model-based SE  \cite{enhancementgan,enhancementflow, enhancementvae,sgmsep,refgerkmannenhancement1,vpidm, Lay2023bbed}, a diffusion process describes gradual transformation of a clean speech sample $x_0$ onto the noisy speech $y$ with additional Gaussian noise using a forward SDE
\begin{equation}
    \label{eq:forwardSDE}
    dx_t = f(x_t,y,t)dt + g(t)dw_t, 
\end{equation}
where $t \in [0,T]$, $w_t$ is a Brownian motion, and $f$ and $g$ are called the drift and the diffusion coefficients, respectively. 
The reverse SDE governing the reverse process that transforms $x_T$ following $\mathcal{N}(y, \sigma_T^2 \mathbf{I})$ 
 onto $x_0$ is given by \cite{song2021scorebased}
\begin{equation}
    \label{eq:reverseSDE}
        \resizebox{\columnwidth}{!}{$\begin{aligned}
            dx_t = \Big[ f(x_t,y, t) - g(t)^2 \nabla_{x_t} \log p_t(x_t | y) \Big] dt + g(t) d \bar{w}_t
        \end{aligned}
        $}
\end{equation}
where $\nabla_{x_t} \log p_t(x_t | y)$ called a score function is the gradient of log for probability density function (pdf) of $x_t$ given $y$, and $\bar{w}_t$ is a reverse Brownian motion.
The score function needed to evaluate the SDE is approximated by a DNN called a score model $s_\theta (x_t, y,t)$. 
The score model is trained using the denoising score matching (DSM) loss \cite{song2021scorebased}, $\mathcal{L}_{DSM}$, which is defined as
\begin{equation}
    \label{eq:dsmloss}           
            \mathcal{L}_{DSM}:= 
            \mathbf{E} \Big\lVert s_{\theta}(x_t,y,t) 
            -\nabla_{x_t} \log p_t (x_t |x_0,y) \Big\rVert^2
\end{equation}
where $t$ is randomly chosen from $U[0,T]$, a uniform distribution between $0$ and $T$, and $x_t$ is sampled from a distribution $p_{t}(x_t|x_0,y)$ called the perturbation kernel which is a Gaussian distribution determined by assumed $f$ and $g$. 
It was reported
\cite{enhancementgan,enhancementflow, enhancementvae,sgmsep,refgerkmannenhancement1,vpidm,Lay2023bbed } that NFEs greater than 25 were required to achieve their performances for those diffusion model-based SE models.

\subsection{Flow Matching-based Speech Enhancement (FlowSE)}
 FlowSE \cite{lee2025flowse} based on FM \cite{lipmanflow, tong2024improving} models a CNF which transforms a random vector following $p_1(x_1 |y):=\mathcal{N}(x_1 | y, \sigma^2 \mathbf{I})$, where $\sigma\geq 0$ is a hyperparameter, into a clean speech distribution given a noisy speech $y$ $q(x_0|y)$ described by an ODE:
 \begin{equation}
 \label{eq:flowseode}
     \frac{d \psi_t(x_1|y)}{dt} := v_t(\psi_t(x_1|y)|y), \psi_1(x_1 |y ) = x_1,
 \end{equation}
 where $\psi_t(x_1 | y)$ and $v_t(x_t | y)$ conditioned on $y$ are called a flow and a vector field, respectively, and $x_1 \sim p_1(x_1|y)$.
 The goal of this formulation is to find the vector field or the flow such that $x_t: =\psi_t(x_1 | y)$ has a pdf $p_t(x_t|y)$ satisfying $p_0(x_0 | y) = q(x_0 | y)$. 
 It is noted that the time index $t$ in (\ref{eq:flowseode}) aligns with the diffusion model in the subsection 2.1, which is not same as that in \cite{lee2025flowse}. 
 FlowSE employed the modified OT conditional vector field with a conditional probability path $p_t(x_t|x_0,y)=\mathcal{N}(x_t| \mu_t (x_0,y), \sigma_t^2 \mathbf{I})$ in which 
 \begin{equation}
     \label{eq:flowsemeanstd}
     \mu_t(x_0,y)=(1-t)x_0 + t y,  \quad \sigma_t = t \sigma.
 \end{equation}
 Then, the target vector field $v_t(x_t|x_0,y)$ becomes 
 \begin{equation}
     \label{eq:targetvectorfield}
      v_t(x_t | x_0,y) = \frac{\frac{d}{dt}\sigma_t}{\sigma_t}(x_t-\mu_t(x_0,y)) + \frac{d}{dt} \mu_t (x_0,y). 
  \end{equation}
The vector field model $v_\theta(x_t,y,t)$ is trained with the CFM loss $\mathcal{L}_{CFM}$ given by
 \begin{equation}
     \label{CFMlossflowse}
     \mathcal{L}_{CFM}:=\mathbf{E} \lVert v_\theta (x_t ,y,t) - v_t(x_t|x_0,y)\rVert^2,
 \end{equation}
 where $t$ is from  $U[t_\delta, 1]$ with $0<t_\delta < 1$ and $x_t$ is from the conditional probability path $p_t(x_t|x_0,y)$. 
 
 In the inference phase, $v_\theta(x_t, y, t)$ is numerically integrated starting with $x_1$ sampled from $p_1(x_1 | y)$. 
 Euler method is adopted as the numerical integrator in \cite{lee2025flowse}.
 Given $N$ time points $t_0=0<t_1=t_\delta<t_2<...<t_N=1$ in $[0,1]$, the clean speech estimate $x_{0}$ is generated from
 \begin{equation}
     x_{t_{i-1}}=x_{t_{i}}+(t_{i-1}-t_{i})v_{\theta}(x_{t_i},y,t_i).
 \end{equation}
 FlowSE with the NFE of 5 achieved performance comparable to a diffusion-based model \cite{Lay2023bbed} with the NFE of 60 and the fine-tuning method for the diffusion-based model \cite{lay2024singlesgmsecrp} with the NFE of 5 \cite{lee2025flowse}. 
It was also shown that FlowSE can be interpreted as a diffusion model-based SE model with a specific SDE \cite{lee2025flowse}.

\subsection{Diffusion-based Stochastic Regenration Model for SE (StoRM)}
StoRM \cite{lemercier2023storm} consists of a predictive model and a generative model based on a diffusion model. 
The predictive model denoted as $D_\phi$ estimates a clean speech $x_0$ from noisy speech $y$, and the estimated speech $D_\phi(y)$ is used as an additional input to the score model $s_\theta(x_t,y,D_\phi(y),t)$ and also utilized to sample the starting point of the reverse process. 
The loss function to train these models is the weighted summation of the mean squared error (MSE) loss $\mathcal{L}_{1}$ for $D_\phi$ and the DSM loss $\mathcal{L}_{2}$ for the score model, i.e., 
\begin{equation}
    \label{eq:stormloss}
    \mathcal{L}^{StoRM}=\alpha\mathcal {L}_{1} +  \mathcal{L}_{2},
\end{equation}
where $\alpha>0$ is a hyperparameter and
\begin{equation}
\label{eq:mseloss}    \mathcal{L}_{1}  := \mathbf{E}\lVert D_\phi(y)-x_0\rVert^2,
\end{equation}
\begin{equation} 
    \mathcal{L}_{2}:=\mathbf{E}\lVert s_\theta (x_t,y,D_\phi(y), t)-\nabla_{x_t} \log p_t (x_t|x_0, D_\phi(y)) \rVert^2, 
       \label{eq:dsmstorm}
\end{equation}
with $t$ from $\mathcal{U}[0,T]$ and $x_t$ sampled from a perturbation kernel $p_t(x_t|x_0,D_\phi(y))$.

During the inference stage of StoRM,  $D_\phi(y)$ is evaluated first and then a clean speech is estimated by integrating the reverse SDE (\ref{eq:reverseSDE}) starting from $\mathcal{N}(D_\phi(y), \sigma_T^2 I)$ using the score model $s_\theta(x_t,y, D_\phi(y), t)$ with conditioning variables $y$ and $D_\phi(y)$. 
StoRM showed better performance than the early diffusion-based model \cite{sgmsep} with only noisy speech as a conditioning variable \cite{lemercier2023storm}. 

\section{Cascading Two Flows for Speech Enhancement (CTFSE)}
We design the first flow to transform a random vector $x_1$ following $p_1(x_1|y)=\mathcal{N}(x_1|y,\sigma^2 \mathbf{I})$ into a crude estimate of clean speech, $D_\theta(x_1,y)$, which is ideally distributed according to $q(x_0|y)$. 
And then, the second flow starts with a sample drawn from a Gaussian distribution centered on $D_\theta(x_1,y)$, $\mathcal{N}(D_\theta(x_1,y), \sigma^2 \mathbf{I})$, and then transforms it using the ODE in (\ref{eq:flowseode}) using the trained vector field model conditioned by both $y$ and $D_\theta(x_1,y)$ to finally obtain $\tilde{x}_0$. 
We use a single vector field model $v_\theta$ for both of the flows by configuring the conditioning variable for the second flow to be a simple summation of $y$ and $D_\theta(x_1,y)$, which is proven to work as a fusion method in many researches \cite{byun2021monaural, li2021two, twostage}.  
The loss for the first flow is the CFM loss in (\ref{CFMlossflowse}) :
\begin{equation}
    \label{eq:firstCTF}
    \mathcal{L}_{1}:=\mathbf{E} \lVert v_\theta (x_t ,y,t) - v_t(x_t|x_0,y)\rVert^2, 
\end{equation}
where $t$ is from $\mathcal{U}[t_\delta, 1]$ and $x_t$ follows $p_t(x_t |x_0,y)$.
To control the computational complexity when two flows are adopted, we fix the number of time steps for the first flow to 1.
Then, $D_\theta(x_1,y)$ is obtained by a simple equation if the Euler method is used for one time step:
\begin{equation}
\label{eq:D_theta_1}
D_\theta(x_1,y)=x_1 -v_\theta(x_1,y, 1).
\end{equation}
The loss for the second flow is again the CFM loss, with different conditioning variables in the vector field model and the target conditional vector field:
\begin{align}
    &\mathcal{L}_{2} :=\mathbf{E}
    \Big\lVert 
   v_\theta  \left(\tilde{x}_{t}, \frac{D_\theta(x_1,y)+y}{2}, t\right)- v_t\left(\tilde{x}_t|x_0,D_\theta(x_1,y)\right)
    \Big\rVert^2,
        \label{eq:secondCTFloss}    
\end{align}
where $t$ is from $\mathcal{U}[t_\delta, 1]$ and $\tilde{x}_t$ is sampled from $p_t(\tilde{x}_t |x_0 , D_\theta (x_1,y))$. 
Additionally, we use the CFM loss $\mathcal{L}_3$ when $t$ is fixed to 1, to enforce $D_\theta(x_1,y)$ to be close to $x_0$:
\begin{equation}
    \label{eq:thirdCTFloss}
    \mathcal{L}_{3}:=\mathbf{E}\lVert v_\theta (x_1 ,y,1) -v_1 (x_1  | x_0, y)\rVert^2,
\end{equation}
where $x_1$ follows $p_1(x_1|x_0,y)$.
It can be shown that $\mathcal{L}_3$ becomes the MSE loss between $D_\theta(x_1,y)$ and a clean speech $x_0$. 
The total loss $\mathcal{L}_{CTF}$ for training is given as a weighted summation of $\mathcal{L}_1, \mathcal{L}_2$ and $\mathcal{L}_3$ 
\begin{equation}
    \label{eq:DFMtotalloss}
    \mathcal{L}_{CTF} := \lambda_{1} \mathcal{L}_{1}  + \lambda_{2}\mathcal{L}_{2} +\lambda_3 \mathcal{L}_{3} 
\end{equation}
where $\lambda_1, \lambda_2, \lambda_3  \geq 0 $ are hyperparameters. 
It is noted that we use the identical model $v_\theta$ to produce $D_\theta (x_1,y)$ and to model the second flow, while StoRM used two different models. 

In the inference phase of CTFSE, $x_1$ is sampled from $p_1 (x_1 | y)$ first and using the Euler method with only one time point, $D_\theta(x_1,y)$ is generated by (\ref{eq:D_theta_1}). 
Given $N$ time points $t_0 =0 < t_1 = t_\delta < t_2 <...<t_N = 1$, a clean speech estimate $\tilde{x}_0 $ is generated by the Euler method
\begin{equation}
\begin{aligned}
\label{eq:CTFsampling}
    \tilde{x}_{t_N} &\sim \mathcal{N}(D_\theta(x_1,y), \sigma^2 \mathbf{I})\\ \tilde{x}_{t_{i-1}}&= \tilde{x}_{t_i}+(t_{i-1}-t_i) v_\theta \left(\tilde{x}_{t_i}, \frac{D_\theta(x_1,y)+y}{2}, t_i \right).
\end{aligned}
\end{equation}

\section{Experimental settings}\begin{figure}[!t]
    \centering
    \includegraphics[width=\columnwidth]{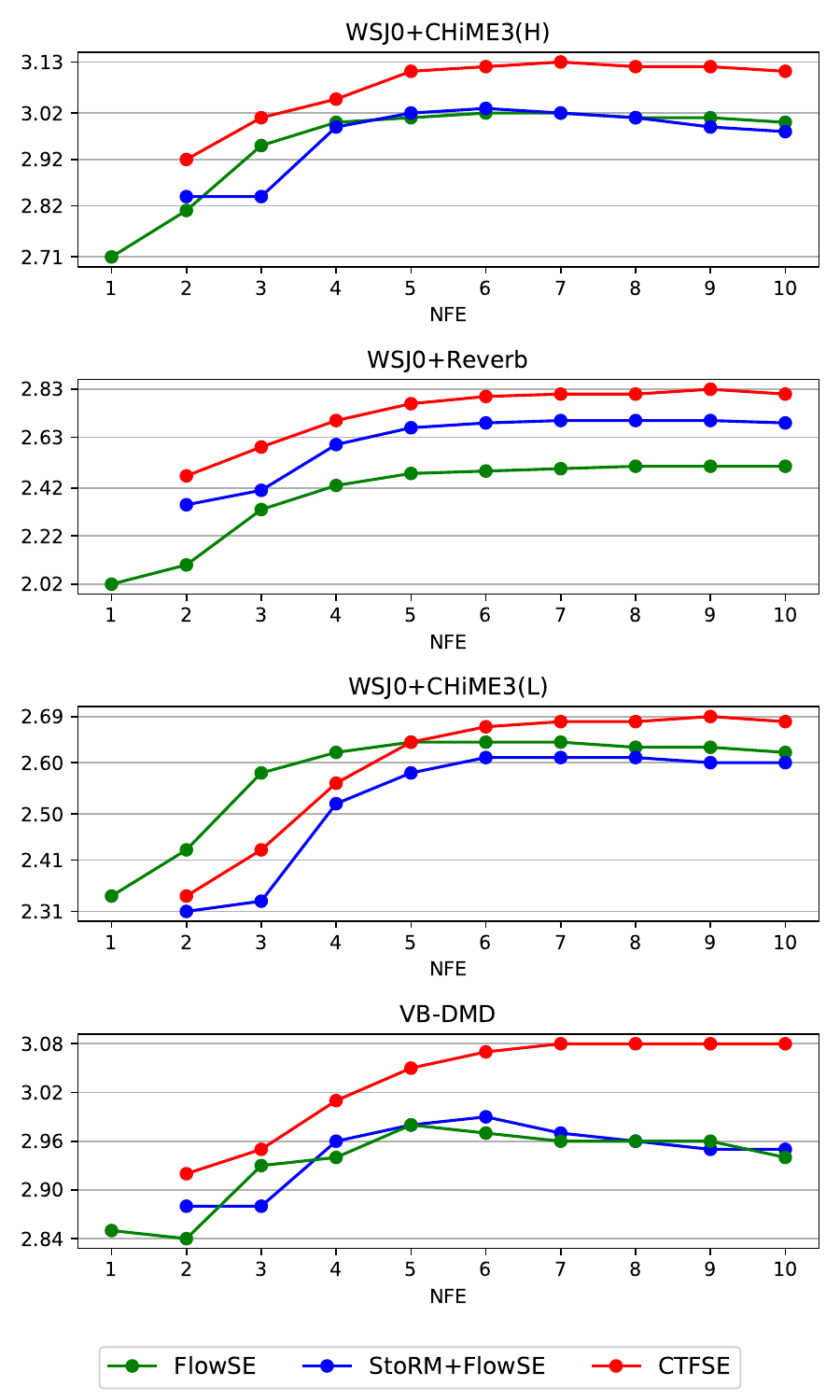} 
    \caption{Comparison of WB-PESQ scores as a function of the NFE for FlowSE, StoRM+FlowSE, and CTFSE.}
    \label{fig:wb_pesq_scores}
    \vspace{-3mm}
\end{figure}
We evaluated the performance of the proposed and compared methods for two versions of WSJ0+CHiME3 dataset, WSJ0+Reverb dataset, and VoiceBank-DEMAND (VB-DMD) dataset. 
WSJ0+CHiME3 (H) and WSJ0+CHiME (L) datasets were constructed by mixing clean speech utterances from the Wall Street Journal (WSJ0) dataset \cite{garofolo1993csriwsj0} and environmental noises from the CHiME3 \cite{barker2015chime} dataset with the signal-to-noise ratio (SNR) between 0 and 20 dB for WSJ0+CHiME3 (H), and -4 and 16 dB for WSJ0+CHiME3 (L). 
WSJ0+Reverb was constructed convolving each utterance from the WSJ0 dataset with a simulated room impulse response (RIR) as in \cite{lemercier2023storm}. 
The dimension of the room was randomly chosen from [5,15], [5,15], and [2,6] m for the length, width, and height, respectively, and the $T_{60}$ was selected in [0.4,1.0] s. 
Then anechoic target speech is generated by simulating the same room with an absorption coefficient of 0.99. 
These three datasets, WSJ0+CHiME3 (H), WSJ0+CHiME3 (L) and WSJ0+Reverb, were created using the source codes\footnote{\texttt{ https://github.com/sp-uhh/storm}} provided by the authors of \cite{lemercier2023storm}.
The VB-DMD dataset \cite{valentini2016investigatingvctk}, which is publicly available, is generated by mixing clean speech from the VCTK dataset \cite{veaux2013voice} with eight real-recorded noise samples from the DEMAND database \cite{thiemann2013diversedemand} and two artificially generated noise samples (babble and speech shaped) at SNRs of 0,5,10, and 15 dB. 
The SNRs for the test set are 2.5, 7.5, 12.5 and 17.5 dB. 

The clean and noisy speech signals, $x_0,y,$ are the magnitude-compressed complex-valued spectrograms in $\mathbb{C}^{K\times F}$ as in \cite{lemercier2023storm}. 
We used a lighter configuration of the NCSN++ architecutre denoted NCSN++M as in \cite{lemercier2023storm, lemercier2023analysing} for the neural network $v_\theta$. 
NCSN++M has 27.8 M parameters, while NCSN++ has 65.0 M parameters. 
We trained the neural network $v_\theta$ using Adam optimizer \cite{adam} with a learning rate of 0.0001 and a batch size of 4.
An exponential moving average with a decay of 0.999 was utilized. 
$t_\delta$ and $\sigma$ were set to 0.03 and 0.5, respectively. 
We set $\lambda_{1}=\lambda_{2} = \lambda_{3} =1$ in (\ref{eq:DFMtotalloss}). 
We trained the model for a maximum of 1,000 epochs with early stopping based on the validation loss with a patience of 50 epochs. 
For the generation process in the subsection 3.2, the time points $0=t_0 < t_1 =t_\delta < t_2 <...<t_N=1$ were chosen so that $t_{i-1}-t_{i}$ has the same value for $i \in \{2,...,N\}$ for $N \geq 2$. 
In the case of $N=1$, we set $t_0=0$ and $ t_1=1$. 

The diffusion model-based SE method SGMSE+M \cite{lemercier2023storm}, StoRM \cite{lemercier2023storm} with a predictive model and a score model, and a flow matching-based model FlowSE \cite{lee2025flowse} were compared with the proposed CTFSE. 
Additionally, we have implemented the StoRM system with a flow matching-based second stage, StoRM+FlowSE, and compared it with other systems. 
All architectures for the neural networks were NCSN++M and only the number of conditioning variables differ from each other. 
For numerical intergration for SDEs, we adopted the predictor-corrector scheme \cite{song2021scorebased} with the Euler-Maruyama method as a predictor and one step of annealed Langevin dynamics correction for SGMSE+M and StoRM, and the Euler method for FlowSE and StoRM+FlowSE for ODEs. 
We utilized the pre-trained SGMSE+M, StoRM models from checkpoints\footnotemark[1] shared by the authors of \cite{lemercier2023storm}, and implemented the remaining baseline models.  

We have evaluated the wideband extension to perceptual evaluation of speech quality (WB-PESQ) scores\cite{widebandpesq},  Deep Noise Supression mean opinion score (DNSMOS) \cite{salas2013subjective}, Extended Short-Time Objective Intelligibility (ESTOI) \cite{jensen2016algorithm}, Scale-Invaraiant Signal-to-Distortion Ratio (SI-SDR) \cite{le2019sdr},  wav2vec MOS (WVMOS) \cite{Andreev_2023}, and DNSMOS P.835 including SIG, BAK and OVRL \cite{reddy2022dnsmos}.

\begin{table*}[!t]
\renewcommand{\arraystretch}{1.15} 
\setlength{\tabcolsep}{5.5pt} 



\caption{Speech enhancement performances for proposed and compared methods on the WSJ0+CHiME3 (H), WSJ0+CHiME3 (L), WSJ0+REVERB and VB-DMD datasets. $^*$ indicates that the results of the model come from checkpoints shared by the authors of \cite{lemercier2023storm}. }
\centering
\label{table:maintable}
\begin{tabular}{cccccccccc}
\hline
\multicolumn{10}{c}{\textbf{Tranined and Tested on WSJ0+CHiME3 (H)}}                                                                                                                                    \\ \hline
\textbf{METHOD} & \textbf{NFE} & \textbf{WB-PESQ}   & \textbf{ESTOI}     & \textbf{SI-SDR}     & \textbf{WVMOS}     & \textbf{DNSMOS}    & \textbf{SIG}       & \textbf{BAK}       & \textbf{OVRL}      \\ \hline
SGMSE+M         & 100          & 2.82±0.04          & 0.92±0.00          & 17.44±0.34          & 3.77±0.02          & 3.93±0.02          & 3.55±0.01          & \textbf{4.19±0.00} & 3.33±0.01          \\
StoRM           & 101          & 2.91±0.04          & 0.92±0.00          & 17.73±0.33          & 3.77±0.03          & 4.00±0.01          & \textbf{3.60±0.01} & 4.12±0.01          & 3.32±0.01          \\
FlowSE          & 6            & 3.02±0.04          & 0.93±0.00          & 18.71±0.34          & 3.86±0.03          & 4.03±0.01          & \textbf{3.60±0.01} & 4.16±0.00          & 3.35±0.01          \\
StoRM+FlowSE    & 6            & 3.03±0.04          & 0.93±0.00          & 18.83±0.34          & 3.86±0.03          & 4.04±0.01          & \textbf{3.60±0.01} & \textbf{4.19±0.00} & \textbf{3.37±0.01} \\
CTFSE           & 6            & \textbf{3.12±0.04} & \textbf{0.94±0.00} & \textbf{19.37±0.33} & \textbf{4.02±0.02} & \textbf{4.05±0.01} & \textbf{3.60±0.01} & \textbf{4.19±0.00} & \textbf{3.37±0.01} \\
CTFSE           & 7            & \textbf{3.13±0.04} & \textbf{0.94±0.00} & 19.20±0.33          & 3.96±0.02          & \textbf{4.05±0.01} & \textbf{3.60±0.01} & \textbf{4.19±0.00} & \textbf{3.37±0.01} \\ \hline
\multicolumn{10}{c}{\textbf{Tranined and Tested on WSJ0+CHiME3 (L)}}                                                                                                                                    \\ \hline
\textbf{METHOD} & \textbf{NFE} & \textbf{PESQ}      & \textbf{ESTOI}     & \textbf{SI-SDR}     & \textbf{WVMOS}     & \textbf{DNSMOS}    & \textbf{SIG}       & \textbf{BAK}       & \textbf{OVRL}      \\ \hline
SGMSE+M$^*$     & 100          & 2.30±0.05          & 0.85±0.01          & 13.19±0.38          & 3.65±0.03          & 3.83±0.02          & 3.51±0.01          & 4.19±0.00          & 3.28±0.01          \\
StoRM$^*$       & 101          & 2.55±0.05          & 0.88±0.01          & 14.91±0.33          & 3.73±0.03          & 4.00±0.01          & 3.57±0.01          & 4.05±0.01          & 3.26±0.01          \\
FlowSE          & 5            & 2.64±0.05          & \textbf{0.89±0.01} & 15.34±0.33          & 3.72±0.03          & 4.01±0.01          & 3.57±0.01          & 4.19±0.00          & 3.34±0.01          \\
StoRM+FlowSE    & 6            & 2.61±0.05          & \textbf{0.89±0.01} & 15.46±0.33          & 3.69±0.03          & 4.02±0.01          & 3.59±0.01          & 4.18±0.00          & 3.35±0.01          \\
CTFSE           & 5            & 2.64±0.05          & \textbf{0.89±0.01} & \textbf{15.96±0.33} & \textbf{3.85±0.03} & 4.01±0.02          & 3.57±0.01          & \textbf{4.20±0.00} & 3.34±0.01          \\
CTFSE           & 9            & \textbf{2.69±0.05} & \textbf{0.89±0.01} & 15.34±0.33 & 3.71±0.03 & \textbf{4.03±0.01} & \textbf{3.60±0.01}          & 4.19±0.00 & \textbf{3.36±0.01}          \\ \hline
\multicolumn{10}{c}{\textbf{Tranined and Tested on WSJ0+Reverb}}                                                                                                                                        \\ \hline
\textbf{METHOD} & \textbf{NFE} & \textbf{PESQ}      & \textbf{ESTOI}     & \textbf{SI-SDR}     & \textbf{WVMOS}     & \textbf{DNSMOS}    & \textbf{SIG}       & \textbf{BAK}       & \textbf{OVRL}      \\ \hline
SGMSE+M$^*$     & 100          & 2.33±0.03          & 0.82±0.01          & -0.21±0.67          & 3.43±0.03          & 3.89±0.02          & 3.20±0.02          & 4.05±0.01          & 2.84±0.02          \\
StoRM$^*$       & 101          & 2.52±0.03          & 0.85±0.00          & 5.54±0.32           & 3.61±0.03          & 3.97±0.01          & \textbf{3.29±0.02} & 4.10±0.01          & \textbf{2.98±0.02} \\
FlowSE          & 9            & 2.51±0.03          & 0.85±0.00          & 4.01±0.35           & 3.59±0.03          & 3.99±0.01          & 3.24±0.02          & \textbf{4.13±0.00} & 2.91±0.02          \\
StoRM+FlowSE    & 9            & 2.70±0.03          & 0.87±0.00          & 6.37±0.29           & 3.74±0.02          & \textbf{4.01±0.01} & 3.25±0.02          & \textbf{4.13±0.00} & 2.92±0.02          \\
CTFSE           & 9            & \textbf{2.83±0.03} & \textbf{0.89±0.00} & \textbf{7.25±0.30}  & \textbf{3.77±0.02} & \textbf{4.01±0.01} & 3.27±0.02          & \textbf{4.13±0.00} & 2.95±0.02          \\ \hline
\multicolumn{10}{c}{\textbf{Tranined and Tested on VB-DMD}}                                                                                                                                             \\ \hline
\textbf{METHOD} & \textbf{NFE} & \textbf{PESQ}      & \textbf{ESTOI}     & \textbf{SI-SDR}     & \textbf{WVMOS}     & \textbf{DNSMOS}    & \textbf{SIG}       & \textbf{BAK}       & \textbf{OVRL}      \\ \hline
SGMSE+M         & 100          & 2.80±0.04          & 0.86±0.01          & 16.19±0.39          & 4.27±0.02          & 3.54±0.02          & 3.48±0.01          & 3.95±0.02          & 3.15±0.02          \\
StoRM$^*$       & 101          & 2.90±0.04          & 0.87±0.01          & 18.48±0.23          & 4.29±0.02          & 3.56±0.02          & \textbf{3.50±0.01} & 4.02±0.01          & \textbf{3.20±0.01} \\
FlowSE          & 5            & 2.98±0.05          & 0.87±0.01          & 18.97±0.23          & \textbf{4.30±0.02} & \textbf{3.58±0.02} & 3.48±0.01          & \textbf{4.05±0.01} & \textbf{3.20±0.01} \\
StoRM+FlowSE    & 6            & 2.99±0.05          & 0.87±0.01          & 18.65±0.24          & 4.26±0.03          & \textbf{3.58±0.02} & 3.49±0.01          & 4.02±0.01          & \textbf{3.20±0.01} \\
CTFSE           & 5            & 3.05±0.05          & \textbf{0.88±0.01} & \textbf{19.13±0.24} & 4.26±0.03          & \textbf{3.58±0.02} & 3.48±0.01          & 4.03±0.01          & \textbf{3.20±0.01} \\
CTFSE           & 7            & \textbf{3.08±0.05} & 0.87±0.01          & 18.97±0.24          & 4.26±0.02          & \textbf{3.58±0.02} & 3.49±0.01          & 4.03±0.01          & \textbf{3.20±0.01}\\ \hline

\end{tabular}

\end{table*}

\section{Results}
Table \ref{table:maintable} summarizes the performances with 95\% confidence intervals and NFEs for the proposed and compared methods on the four datasets, WSJ0+CHiME3 (H), WSJ0+CHiME3 (L), WSJ0+Reverb, and VB-DMD. 
The NFEs for SGMSE+M and StoRM were set to 100 and 101 as in \cite{lemercier2023storm}, and those for FlowSE, StoRM+FlowSE and CTFSE were selected to maximize the average WB-PESQ scores. 
For fair comparison, the performances for the CTFSE with the NFE of 5 or 6 are also shown when the compared methods have lower NFE of 5 or 6. 
On average, the proposed CTFSE showed the best performance for most of the mesures with the NFE less than 10. 
Compared with StoRM+FlowSE which had the same NFE but twice the parameters, the proposed CTFSE exhibited similar or better performances and the performance improvement was bigger for WSJ0+CHiME3 (H) and WSJ0+Reverb. 

Figure \ref{fig:wb_pesq_scores} shows the WB-PESQ scores as a function of NFE for FlowSE, StoRM+FlowSE, and CTFSE. 
We can see that CTFSE outperformed other methods at the same NFE except for the WSJ0+CHiME3 (L) with the NFE less than 5.

\section{Conclusion}
In this work, we proposed a speech enhancement cascading two flows with the same vector field model. 
The first flow produces the crude estimate of clean speech which is used as the mean of the starting point of the second flow and summed up to noisy speech to be used as a conditioning variable for the second flow. 
The loss function to train the vector field model is the weighted combination of the CFM losses for two flows. 
Experimental results demonstrated that the proposed method showed comparable or better performance to the previously proposed diffusion model- or flow matching-based SE methods for four datasets. 
\section{Acklowedgements}
This research was supported by the MSIT (Ministry of Science and ICT), Korea, under the ITRC (Information Technology Research Center) support program (IITP-2025-RS-2021-II211835) supervised by the IITP (Institute of Information Communications Technology Planning Evaluation) and Institute of Information \& Communications Technology Planning \& Evaluation (IITP) grant funded by MSIT (No.2019-0-01842, Artificial Intelligence Graduate School Program (GIST)).
\clearpage

\begingroup
\scriptsize 
\bibliographystyle{IEEEtran}
\bibliography{mybib}

\begin{thebibliography}{10}
\providecommand{\url}[1]{#1}
\csname url@samestyle\endcsname
\providecommand{\newblock}{\relax}
\providecommand{\bibinfo}[2]{#2}
\providecommand{\BIBentrySTDinterwordspacing}{\spaceskip=0pt\relax}
\providecommand{\BIBentryALTinterwordstretchfactor}{4}
\providecommand{\BIBentryALTinterwordspacing}{\spaceskip=\fontdimen2\font plus
\BIBentryALTinterwordstretchfactor\fontdimen3\font minus \fontdimen4\font\relax}
\providecommand{\BIBforeignlanguage}[2]{{%
\expandafter\ifx\csname l@#1\endcsname\relax
\typeout{** WARNING: IEEEtran.bst: No hyphenation pattern has been}%
\typeout{** loaded for the language `#1'. Using the pattern for}%
\typeout{** the default language instead.}%
\else
\language=\csname l@#1\endcsname
\fi
#2}}
\providecommand{\BIBdecl}{\relax}
\BIBdecl

\bibitem{loizou2007speech}
P.~C. Loizou, \emph{Speech enhancement: theory and practice}.\hskip 1em plus 0.5em minus 0.4em\relax CRC press, 2007.

\bibitem{minseungenhancement}
M.~Kim and J.~W. Shin, ``Improved speech enhancement considering speech psd uncertainty,'' \emph{IEEE/ACM Transactions on Audio, Speech, and Language Processing (TASLP)}, vol.~30, pp. 1939--1951, 2022.

\bibitem{seinenhancement}
S.~Cheong, M.~Kim, and J.~W. Shin, ``Postfilter for dual channel speech enhancement using coherence and statistical model-based noise estimation,'' \emph{Sensors}, vol.~24, no.~12, 2024.

\bibitem{enhancementwangtan}
K.~Tan and D.~Wang, ``Learning complex spectral mapping with gated convolutional recurrent networks for monaural speech enhancement,'' \emph{TASLP}, vol.~28, pp. 380--390, 2020.

\bibitem{kim2022factorizedfmvdr}
H.~Kim, K.~Kang, and J.~W. Shin, ``Factorized mvdr deep beamforming for multi-channel speech enhancement,'' \emph{IEEE Signal Processing Letters}, vol.~29, pp. 1898--1902, 2022.

\bibitem{kim2021target}
H.~Kim and J.~W. Shin, ``Target exaggeration for deep learning-based speech enhancement,'' \emph{Digital Signal Processing}, vol. 116, pp. 103--109, 2021.

\bibitem{enhancementgan}
D.~Baby and S.~Verhulst, ``Sergan: Speech enhancement using relativistic generative adversarial networks with gradient penalty,'' in \emph{IEEE International Conference on Acoustics, Speech and Signal Processing (ICASSP)}, 2019, pp. 106--110.

\bibitem{enhancementflow}
A.~A. Nugraha, K.~Sekiguchi, and K.~Yoshii, ``A flow-based deep latent variable model for speech spectrogram modeling and enhancement,'' \emph{TASLP}, vol.~28, pp. 1104--1117, 2020.

\bibitem{enhancementvae}
X.~Bie, S.~Leglaive, X.~Alameda-Pineda, and L.~Girin, ``Unsupervised speech enhancement using dynamical variational autoencoders,'' \emph{TASLP}, vol.~30, pp. 2993--3007, 2022.

\bibitem{sgmsep}
J.~Richter, S.~Welker, J.-M. Lemercier, B.~Lay, and T.~Gerkmann, ``Speech enhancement and dereverberation with diffusion-based generative models,'' \emph{TASLP}, vol.~31, pp. 2351--2364, 2023.

\bibitem{refgerkmannenhancement1}
J.-M. Lemercier, J.~Richter, S.~Welker, and T.~Gerkmann, ``Analysing diffusion-based generative approaches versus discriminative approaches for speech restoration,'' in \emph{ICASSP}, 2023.

\bibitem{vpidm}
Z.~Guo, Q.~Wang, J.~Du, J.~Pan, Q.-F. Liu, and C.-H. Lee, ``A variance-preserving interpolation approach for diffusion models with applications to single channel speech enhancement and recognition,'' \emph{TASLP}, vol.~32, pp. 3025--3038, 2024.

\bibitem{Lay2023bbed}
B.~Lay, S.~Welker, J.~Richter, and T.~Gerkmann, ``Reducing the prior mismatch of stochastic differential equations for diffusion-based speech enhancement,'' in \emph{Interspeech}, 2023.

\bibitem{lemercier2023storm}
J.-M. Lermercier, J.~Richter, S.~Welker, and T.~Gerkmann, ``Storm: A diffusion-based stochastic regeneration model for speech enhancement and dereverberation,'' \emph{TASLP}, 2023.

\bibitem{thunderinterspeech2024}
T.~Trachu, C.~Piansaddhayanon, and E.~Chuangsuwanich, ``Unified regression-diffusion speech enhancement with a single reverse step using brownian bridge,'' in \emph{interpseech}, 2024.

\bibitem{lipmanflow}
Y.~Lipman, R.~T.~Q. Chen, H.~Ben-Hamu, M.~Nickel, and M.~Le, ``Flow matching for generative modeling,'' in \emph{International Conference on Learning Representations (ICLR)}, 2023.

\bibitem{tong2024improving}
A.~Tong, K.~FATRAS, N.~Malkin, G.~Huguet, Y.~Zhang, J.~Rector-Brooks, G.~Wolf, and Y.~Bengio, ``Improving and generalizing flow-based generative models with minibatch optimal transport,'' \emph{Transactions on Machine Learning Research}, 2024.

\bibitem{speechflow}
A.~H. Liu, M.~Le, A.~Vyas, B.~Shi, A.~Tjandra, and W.-N. Hsu, ``Generative pre-training for speech with flow matching,'' in \emph{ICLR}, 2024.

\bibitem{lee2025flowse}
S.~Lee, S.~Cheong, S.~Han, and J.~W. Shin, ``Flowse: Flow matching-based speech enhancement,'' in \emph{ICASSP}, 2025.

\bibitem{flowavse}
C.~Jung, S.~Lee, J.-H. Kim, and J.~S. Chung, ``Flowavse: Efficient audio-visual speech enhancement with conditional flow matching,'' in \emph{Interspeech}, 2024, pp. 2210--2214.

\bibitem{tai2023revisiting}
W.~Tai, F.~Zhou, G.~Trajcevski, and T.~Zhong, ``Revisiting denoising diffusion probabilistic models for speech enhancement: Condition collapse, efficiency and refinement,'' in \emph{AAAI}, vol.~37, no.~11, 2023, pp. 13\,627--13\,635.

\bibitem{serra2022universal}
J.~Serr{\`a}, S.~Pascual, J.~Pons, R.~O. Araz, and D.~Scaini, ``Universal speech enhancement with score-based diffusion,'' \emph{arXiv preprint}, vol. arXiv:2206.03065, 2022.

\bibitem{kim2024guided}
D.~Kim, D.~H. Yang, D.~Kim, J.~H. Chang, J.~Yang, J.~Choi, M.~Lee, and H.~g.~Moon, ``Guided conditioning with predictive network on score-based diffusion model for speech enhancement,'' in \emph{Interspeech}, 2024, pp. 1190--1194.

\bibitem{song2021scorebased}
Y.~Song, J.~Sohl-Dickstein, D.~P. Kingma, A.~Kumar, S.~Ermon, and B.~Poole, ``Score-based generative modeling through stochastic differential equations,'' in \emph{ICLR}, 2021.

\bibitem{lay2024singlesgmsecrp}
B.~Lay, J.-M. Lermercier, J.~Richter, and T.~Gerkmann, ``Single and few-step diffusion for generative speech enhancement,'' in \emph{ICASSP}, 2024, pp. 626--630.

\bibitem{byun2021monaural}
J.~Byun and J.~W. Shin, ``Monaural speech separation using speaker embedding from preliminary separation,'' \emph{TASLP}, vol.~29, pp. 2753--2763, 2021.

\bibitem{li2021two}
A.~Li, W.~Liu, C.~Zheng, C.~Fan, and X.~Li, ``Two heads are better than one: A two-stage complex spectral mapping approach for monaural speech enhancement,'' \emph{TASLP}, vol.~29, 2021.

\bibitem{twostage}
X.~Hao, X.~Su, S.~Wen, Z.~Wang, Y.~Pan, F.~Bao, and W.~Chen, ``Masking and inpainting: A two-stage speech enhancement approach for low snr and non-stationary noise,'' in \emph{ICASSP}, 2020, pp. 6959--6963.

\bibitem{garofolo1993csriwsj0}
J.~S. Garofolo, D.~Graff, D.~Paul, and D.~Pallett, ``Csr-i (wsj0) complete,'' 1993.

\bibitem{barker2015chime}
J.~Barker, R.~Marxer, E.~Vincent, and S.~Watanabe, ``The third ‘chime’ speech separation and recognition challenge: Dataset, task and baselines,'' in \emph{Proceedings of the IEEE Workshop on Automatic Speech Recognition and Understanding (ASRU)}.\hskip 1em plus 0.5em minus 0.4em\relax IEEE, 2015, pp. 504--511.

\bibitem{valentini2016investigatingvctk}
C.~Valentini-Botinhao, X.~Wang, S.~Takaki, and J.~Yamagishi, ``Investigating rnn-based speech enhancement methods for noise-robust text-to-speech.'' in \emph{SSW}, 2016, pp. 146--152.

\bibitem{veaux2013voice}
C.~Veaux, J.~Yamagishi, and S.~King, ``The voice bank corpus: Design, collection and data analysis of a large regional accent speech database,'' in \emph{Proc. Int. Conf. Oriental COCOSDA Held Jointly With Conf. Asian Spoken Lang. Res. Eval. (O-COCOSDA/CASLRE),}.\hskip 1em plus 0.5em minus 0.4em\relax IEEE, 2013, pp. 1--4.

\bibitem{thiemann2013diversedemand}
J.~Thiemann, N.~Ito, and E.~Vincent, ``The diverse environments multi-channel acoustic noise database (demand): A database of multichannel environmental noise recordings,'' in \emph{Proceedings of Meetings on Acoustics}, vol.~19.\hskip 1em plus 0.5em minus 0.4em\relax AIP Publishing, 2013.

\bibitem{lemercier2023analysing}
J.-M. Lemercier, J.~Richter, S.~Welker, and T.~Gerkmann, ``Analysing diffusion-based generative approaches versus discriminative approaches for speech restoration,'' in \emph{ICASSP}, 2023.

\bibitem{adam}
D.~P. Kingma and J.~Ba, ``Adam: A method for stochastic optimization,'' \emph{arXiv preprint arXiv:1412.6980}, 2014.

\bibitem{widebandpesq}
\emph{Wideband extension to recommendation P.862 for the assessment of wideband telephone networks and speech codec}, International Telecommunication Union, Geneva, 2007, iTU-T Recommendation P.862.2.

\bibitem{salas2013subjective}
O.~F. Salas, V.~Adzic, and H.~Kalva, ``Subjective quality evaluations using crowdsourcing,'' in \emph{2013 Picture Coding Symposium (PCS)}.\hskip 1em plus 0.5em minus 0.4em\relax IEEE, 2013, pp. 418--421.

\bibitem{jensen2016algorithm}
J.~Jensen and C.~H. Taal, ``An algorithm for predicting the intelligibility of speech masked by modulated noise maskers,'' \emph{TASLP}, vol.~24, no.~11, pp. 2009--2022, 2016.

\bibitem{le2019sdr}
J.~Le~Roux, S.~Wisdom, H.~Erdogan, and J.~R. Hershey, ``Sdr--half-baked or well done?'' in \emph{ICASSP}, 2019, pp. 626--630.

\bibitem{Andreev_2023}
P.~Andreev, A.~Alanov, O.~Ivanov, and D.~Vetrov, ``Hifi++: A unified framework for bandwidth extension and speech enhancement,'' in \emph{ICASSP}, 2023.

\bibitem{reddy2022dnsmos}
C.~K. Reddy, V.~Gopal, and R.~Cutler, ``Dnsmos p.835: A non-intrusive perceptual objective speech quality metric to evaluate noise suppressors,'' in \emph{ICASSP}, 2022, pp. 886--890.

\end{thebibliography}
\endgroup

\end{document}